\documentclass[a4paper]{article}

\usepackage{INTERSPEECH2020}
\usepackage{color}
\usepackage{multirow}
\usepackage{makecell}

\title{Conformer: Convolution-augmented Transformer for Speech Recognition}
\name{Anmol Gulati, James Qin, Chung-Cheng Chiu, Niki Parmar, Yu Zhang, Jiahui Yu, Wei Han, Shibo Wang, Zhengdong Zhang, Yonghui Wu, Ruoming Pang}
\address{Google Inc.}
\email{\{anmolgulati, jamesqin, chungchengc, nikip, ngyuzh, jiahuiyu, weihan, shibow, zhangzd, yonghui, rpang\}@google.com}

\newcommand{\netname}{Conformer}

\usepackage{hyperref}
\begin{document}

\maketitle
\begin{abstract}
Recently Transformer and Convolution neural network (CNN) based models have shown promising results in Automatic Speech Recognition (ASR), outperforming Recurrent neural networks (RNNs). Transformer models are good at capturing content-based global interactions, while CNNs exploit local features effectively.
In this work, we achieve the best of both worlds by studying how to combine convolution neural networks and transformers to model both local and global dependencies of an audio sequence in a parameter-efficient way. To this regard, we propose the convolution-augmented transformer for speech recognition, named \textit{Conformer}. \textit{Conformer} significantly outperforms the previous Transformer and CNN based models achieving state-of-the-art accuracies. On the widely used LibriSpeech benchmark, our model achieves WER of 2.1\%/4.3\% without using a language model and 1.9\%/3.9\% with an external language model on test/testother. We also observe competitive performance of 2.7\%/6.3\% with a small model of only 10M parameters.

\end{abstract}

\section{Introduction}
End-to-end automatic speech recognition (ASR) systems based on neural networks have seen large improvements in recent years.
Recurrent neural networks (RNNs) have been the de-facto choice for ASR~\cite{chiu2018state, rao2017exploring, Ryan19, tara2020} as they can model the temporal dependencies in the audio sequences effectively \cite{graves2012sequence}. Recently, the Transformer architecture based on self-attention~\cite{vaswani2017attention, zhang2020transformer} has enjoyed widespread adoption for modeling sequences due to its ability to capture long distance interactions and the high training efficiency. Alternatively, convolutions have also been successful for ASR~\cite{li2019jasper, kriman2019quartznet, han2020contextnet, sainath2013deep, abdel2014convolutional}, which capture local context progressively via a local receptive field layer by layer.

However, models with self-attention or convolutions each has its limitations. While Transformers are good at modeling long-range global context, they are less capable to extract fine-grained local feature patterns. Convolution neural networks (CNNs), on the other hand, exploit local information and are used as the de-facto computational block in vision. They learn shared position-based kernels over a local window which maintain translation equivariance and are able to capture features like edges and shapes. One limitation of using local connectivity is that you need many more layers or parameters to capture global information. To combat this issue, contemporary work ContextNet~\cite{han2020contextnet} adopts the squeeze-and-excitation module~\cite{hu2018squeeze} in each residual block to capture longer context. However, it is still limited in capturing dynamic global context as it only applies a global averaging over the entire sequence.

Recent works have shown that combining convolution and self-attention improves over using them individually~\cite{bello2019attention}. Together, they are able to learn both position-wise local features, and use content-based global interactions. Concurrently, papers like \cite{yang2019convolutional, yu2018qanet} have augmented self-attention with relative position based information that maintains equivariance. Wu et al. \cite{wu2020lite} proposed a multi-branch architecture with splitting the input into two branches: self-attention and convolution; and concatenating their outputs. Their work targeted mobile applications and showed improvements in machine translation tasks.

\begin{figure}[t]
    \includegraphics[width=0.8\columnwidth]{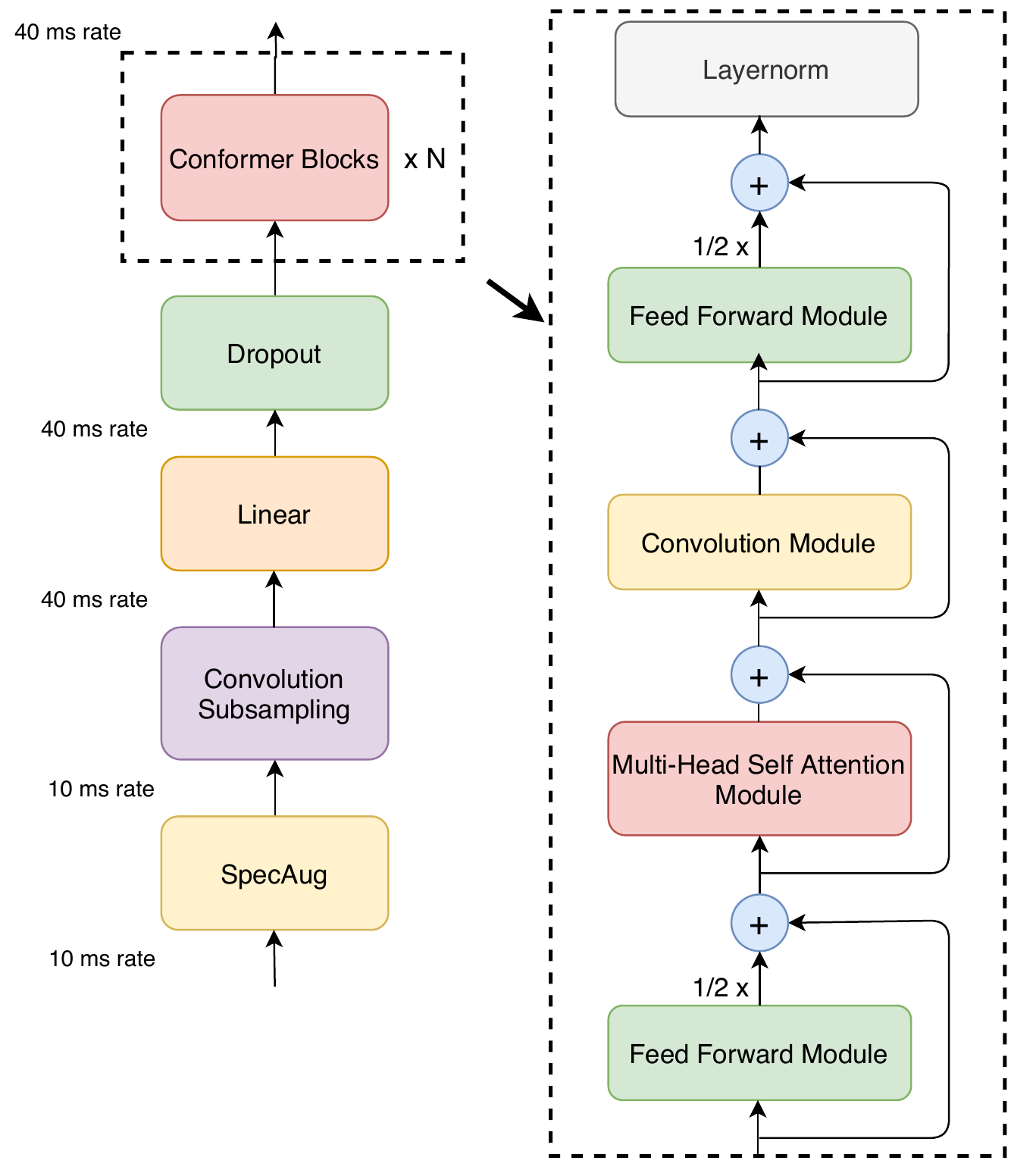}
    \caption{\textbf{Conformer encoder model architecture.} Conformer comprises of two macaron-like feed-forward layers with half-step residual connections sandwiching the multi-headed self-attention and convolution modules. This is followed by a post layernorm.}
    \label{fig:conformer}
\end{figure}

\label{sec:model:convolution}
\begin{figure*}
    \centering
    \includegraphics[width=0.7\textwidth]{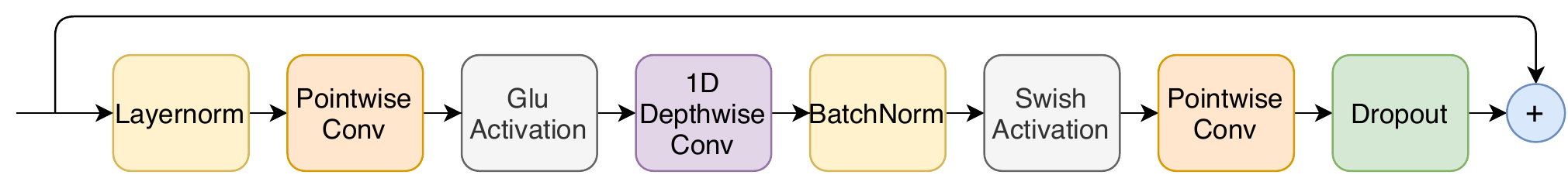}
    \caption{\textbf{Convolution module.} The convolution module contains a pointwise convolution with an expansion factor of 2 projecting the number of channels with a GLU activation layer, followed by a 1-D Depthwise convolution. The 1-D depthwise conv is followed by a Batchnorm and then a swish activation layer.  }
    \label{fig:conv}
\end{figure*}

In this work, we study how to organically combine convolutions with self-attention in ASR models. We hypothesize that both global and local interactions are important for being parameter efficient. To achieve this, we propose a novel combination of self-attention and convolution will achieve the best of both worlds -- self-attention learns the global interaction whilst the convolutions efficiently capture the relative-offset-based local correlations.
Inspired by Wu et al. \cite{wu2020lite, lu2019understanding}, we introduce a novel combination of self-attention and convolution, sandwiched between a pair feed forward modules,  as illustrated in Fig~\ref{fig:conformer}.  

Our proposed model, named Conformer, achieves state-of-the-art results on LibriSpeech, outperforming the previous best published Transformer Transducer \cite{zhang2020transformer} by 15\% relative improvement on the testother dataset with an external language model. We present three models based on model parameter limit constraints of 10M , 30M and 118M.
Our 10M model shows an improvement when compared to similar sized contemporary work \cite{han2020contextnet} with 2.7\%/6.3\% on test/testother datasets. 
Our medium 30M parameters-sized model already outperforms  transformer transducer published in  \cite{zhang2020transformer} which uses 139M model parameters.
With the big 118M parameter model, we are able to achieve 2.1\%/4.3\% without using language models and 1.9\%/3.9\% with an external language model.

We further carefully study the effects of the number of attention heads, convolution kernel sizes, activation functions, placement of feed-forward layers, and different strategies of adding convolution modules to a Transformer-based network, and shed light on how each contributes to the accuracy improvements.
\section{Conformer Encoder}

Our audio encoder first processes the input with a convolution subsampling layer and then with a number of conformer blocks, as illustrated in Figure~\ref{fig:conformer}. The distinctive feature of our model is the use of Conformer blocks in the place of Transformer blocks as in~\cite{zhang2020transformer, karita2019comparative}.

A conformer block is composed of four modules stacked together, i.e, a feed-forward module, a self-attention module, a convolution module, and a second feed-forward module in the end. 
Sections~\ref{sec:model:atten}, \ref{sec:model:convolution}, and \ref{sec:model:ffn} introduce the self-attention,  convolution, and feed-forward modules, respectively. Finally, \ref{sec:model:conformer} describes how these sub blocks are combined.

\subsection{Multi-Headed Self-Attention Module}
\label{sec:model:atten}

We employ multi-headed self-attention (MHSA) while integrating an important technique from Transformer-XL \cite{dai2019transformerxl}, the relative sinusoidal positional encoding scheme. The relative positional encoding allows the self-attention module to generalize better on different input length and the resulting encoder is more robust to the variance of the utterance length. We use pre-norm residual units~\cite{wang-etal-2019-learning-deep, nguyen2019transformers} with dropout which helps training and regularizing deeper models. Figure~\ref{fig:mhsa} below illustrates the multi-headed self-attention block.

\label{sec:model:atten:selfatten}
\begin{figure}[h!]
\includegraphics[width=0.9\columnwidth]{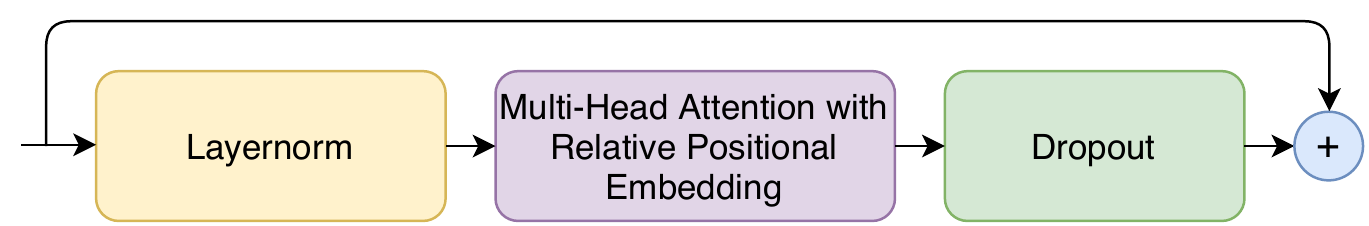}
    \caption{\textbf{Multi-Headed self-attention module.} We use multi-headed self-attention with relative positional embedding in a pre-norm residual unit.} 
    \label{fig:mhsa}
\end{figure}

\subsection{Convolution Module}
Inspired by \cite{wu2020lite}, the convolution module starts with a gating mechanism \cite{dauphin2017language}---a pointwise convolution and a gated linear unit (GLU). This is followed by a single 1-D depthwise convolution layer.
Batchnorm is deployed just after the convolution to aid training deep models. Figure~\ref{fig:conv} illustrates the convolution block.

\subsection{Feed Forward Module}
\label{sec:model:ffn}
The Transformer architecture as proposed in \cite{vaswani2017attention} deploys a feed forward module after the MHSA layer and is composed of two linear transformations and a nonlinear activation in between. A residual connection is added over the feed-forward layers, followed by layer normalization.
This structure is also adopted by Transformer ASR models~\cite{zhang2020transformer, dong2018speech}.

\begin{figure*}
    \centering
    \includegraphics[width=0.5\textwidth]{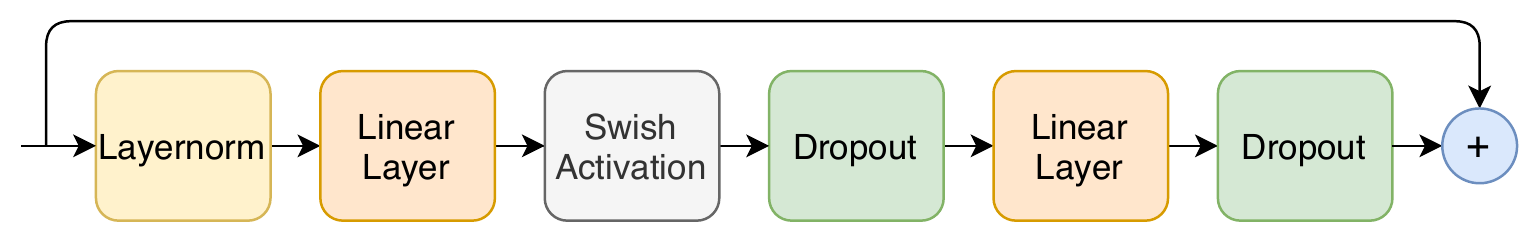}
    \caption{\textbf{Feed forward module.} The first linear layer uses an expansion factor of 4 and the second linear layer projects it back to the model dimension. We use swish activation and a pre-norm residual units in feed forward module.}
    \label{fig:ffn}
\end{figure*}

We follow pre-norm residual units~\cite{wang-etal-2019-learning-deep, nguyen2019transformers} and apply layer normalization within the residual unit and on the input before the first linear layer. We also apply Swish activation \cite{ramachandran2017searching} and dropout, which helps regularizing the network. Figure~\ref{fig:ffn} illustrates the Feed Forward (FFN) module.

\subsection{Conformer Block}
\label{sec:model:conformer}
Our proposed Conformer block contains two Feed Forward modules sandwiching the Multi-Headed Self-Attention module and the Convolution module, as shown in Figure~\ref{fig:conformer}.

This sandwich structure is inspired by Macaron-Net~\cite{lu2019understanding}, which proposes  replacing the original feed-forward layer in the Transformer block into two half-step feed-forward layers, one before the attention layer and one after.  
As in Macron-Net, we employ half-step residual weights in our feed-forward (FFN) modules. The second feed-forward module is followed by a final layernorm layer.
Mathematically, this means, for input $x_{i}$ to a Conformer block $i$, the output $y_{i}$ of the block is:
\begin{equation}
\begin{split}
    \tilde{x_{i}}&= x_{i} + \frac{1}{2}\mathrm{FFN}(x_{i})    \\
    x'_{i}&= \tilde{x_{i}} + \mathrm{MHSA}(\tilde{x_{i}})    \\
    x''_{i}&= x'_{i} + \mathrm{Conv}(x'_{i})    \\
    y_{i}&= \mathrm{Layernorm}(x''_{i} + \frac{1}{2}\mathrm{FFN}(x''_{i}))    \\
\end{split}
\label{model:conformer:equation}
\end{equation}
where FFN refers to the Feed forward module, MHSA refers to the Multi-Head Self-Attention module, and Conv refers to the Convolution module as described in the preceding sections.

Our ablation study discussed in Sec~\ref{ablate_conformer_macaron} compares the Macaron-style half-step FFNs with the vanilla FFN as used in previous works. We find that having two Macaron-net style feed-forward layers with half-step residual connections sandwiching the attention and convolution modules in between provides a significant improvement over having a single feed-forward module in our Conformer architecture.

The combination of convolution and self-attention has been studied before and one can imagine many ways to achieve that. 
Different options of augmenting convolutions with self-attention are studied in Sec~\ref{ablate_conformer_convolution}.
We found that convolution module stacked after the self-attention module works best for speech recognition.
\section{Experiments}
\vskip -0.1in
\label{sec:experiment}
\subsection{Data}
We evaluate the proposed model on the LibriSpeech~\cite{panayotov2015librispeech} dataset, which consists of 970 hours of labeled speech and an additional 800M word token text-only corpus for building language model. We extracted 80-channel filterbanks features computed from a 25ms window with a stride of 10ms. We use SpecAugment \cite{park2019specaugment, largespecaugment} with mask parameter ($F=27$), and ten time masks with maximum time-mask ratio ($p_S = 0.05$), where the maximum-size of the time mask is set to $p_S$ times the length of the utterance.

\subsection{Conformer Transducer}

We identify three models, small, medium and large, with 10M, 30M, and 118M params, respectively, by sweeping different combinations of network depth, model dimensions, number of attention heads and choosing the best performing one within model parameter size constraints. We use a single-LSTM-layer decoder in all our models.
Table ~\ref{tab:model_sizes} describes their architecture hyper-parameters. 

For regularization, we apply dropout \cite{JMLR:v15:srivastava14a} in each residual unit of the conformer, i.e, to the output of each module, before it is added to the module input. We use a rate of $P_{drop} = 0.1$. Variational noise \cite{graves2012sequence, jim1996analysis} is introduced to the model as a regularization. A $\ell_2$ regularization with $1e-6$ weight is also added to all the trainable weights in the network. We train the models with the Adam optimizer \cite{kingma2014adam} with $\beta_{1} = 0.9$, $\beta_{2} = 0.98$ and $\epsilon = 10^{-9}$ and a transformer learning rate schedule \cite{vaswani2017attention}, with 10k warm-up steps and peak learning rate $0.05/\sqrt{\bf{d}}$ where $\bf{d}$ is the model dimension in conformer encoder.   

We use a 3-layer LSTM language model (LM) with width 4096 trained on the LibriSpeech langauge model corpus with the LibriSpeech960h transcripts added, tokenized with the 1k WPM built from LibriSpeech 960h.  The LM has word-level perplexity 63.9 on the dev-set transcripts. The LM weight $\lambda$ for shallow fusion is tuned on the dev-set via grid search. All models are implemented with Lingvo toolkit~\cite{lingvo}.

\begin{table}
\begin{centering}
\caption{Model hyper-parameters for Conformer S, M, and L models, found via sweeping different combinations and choosing the best performing models within the parameter limits.}
\label{tab:model_sizes}
\begin{tabular}{@{}l|ccc@{}}
\hline 
\small
\makecell{Model} & \makecell{Conformer\\ (S)} & \makecell{Conformer\\ (M)} & \makecell{Conformer\\ (L)} \\
\hline 
\hline 
{Num Params (M)} & 10.3 & 30.7 & 118.8  \tabularnewline
{Encoder Layers} & 16 & 16 & 17 \tabularnewline
{Encoder Dim} & 144 & 256 & 512  \tabularnewline
{Attention Heads} & 4 & 4 & 8 \tabularnewline
{Conv Kernel Size} & 32 & 32 & 32 \tabularnewline
{Decoder Layers} & 1 & 1 & 1 \tabularnewline
{Decoder Dim} & 320 & 640 & 640 \tabularnewline
\hline
\end{tabular}
\par\end{centering}
\end{table}

\subsection{Results on LibriSpeech}
\begin{table}[h!]
\caption{Comparison of {\netname} with recent published models. Our model shows improvements consistently over various model parameter size constraints. At 10.3M parameters, our model is 0.7\% better on testother when compared to contemporary work, ContextNet(S) \cite{han2020contextnet}. At 30.7M model parameters our model already significantly outperforms the previous published state of the art results of Transformer Transducer \cite{zhang2020transformer} with 139M parameters.}
\label{tab:results:libri_main_results}
  
  \centering
  \small
  \resizebox{\columnwidth}{!}{%
  \begin{tabular}{@{}lccccc@{}}
    \toprule
    \bfseries Method & \#{\bfseries Params (M)} & \multicolumn{2}{c}{\bfseries WER Without LM} & \multicolumn{2}{c}{\bfseries WER With LM} \\
    \cmidrule(r){3-4} \cmidrule(r){5-6}
     & & \bfseries testclean & \bfseries testother & \bfseries testclean & \bfseries testother \\
    \midrule
    \bfseries Hybrid \\
    \quad Transformer~\cite{wang2019transformer} & - 
    & - & - & 2.26 & 4.85 \\
    \bfseries CTC \\
    \quad QuartzNet ~\cite{kriman2019quartznet} & 19
    &3.90 & 11.28 & 2.69 & 7.25\\
    \bfseries LAS \\
    \quad  Transformer \cite{synnaeve2019endtoend} & 270
    & 2.89 & 6.98 & 2.33 & 5.17 \\
    \quad Transformer \cite{karita2019comparative} & -
    & 2.2 & 5.6 & 2.6 & 5.7 \\
    \quad LSTM & 360 & 2.6 & 6.0 & 2.2 & 5.2 \\
    \bfseries Transducer \\
    \quad Transformer \cite{zhang2020transformer} & 139
    & 2.4 & 5.6 & 2.0 & 4.6 \\
    \quad ContextNet(S) \cite{han2020contextnet} & 10.8
    & 2.9 & 7.0 & 2.3 & 5.5\\
    \quad ContextNet(M) \cite{han2020contextnet} & 31.4
    & 2.4 & 5.4 & \textbf{2.0} & 4.5\\
    \quad ContextNet(L) \cite{han2020contextnet} & 112.7
    & \textbf{2.1} & 4.6 & \textbf{1.9} & 4.1\\
    \midrule
    \bfseries Conformer (Ours)\\
    \quad\netname(S) & 10.3 & \textbf{2.7} & \textbf{6.3} & \textbf{2.1} & \textbf{5.0} \\
    \quad\netname(M) & 30.7 & \textbf{2.3} & \textbf{5.0} & \textbf{2.0} & \textbf{4.3} \\
    \quad\netname(L) & 118.8 & \textbf{2.1} & \textbf{4.3} & \textbf{1.9} & \textbf{3.9} \\
    \bottomrule
  \end{tabular}}
\end{table}

Table~\ref{tab:results:libri_main_results} compares the (WER) result of our model on LibriSpeech test-clean/test-other with a few state-of-the-art models include: ContextNet~\cite{han2020contextnet}, Transformer transducer~\cite{zhang2020transformer}, and QuartzNet~\cite{kriman2019quartznet}.
All our evaluation results round up to 1 digit after decimal point.

Without a language model, the performance of our medium model already achieve competitive results of $2.3$/$5.0$ on test/testother outperforming the best known Transformer, LSTM based model, or a similar sized convolution model. With the language model added, our model achieves the lowest word error rate among all the existing models. This clearly demonstrates the effectiveness of combining Transformer and convolution in a single neural network.

\subsection{Ablation Studies}

\subsubsection{Conformer Block vs. Transformer Block}
\label{ablate_conformer_transformer}
A Conformer block differs from a Transformer block in a number of ways, in particular, the inclusion of a convolution block and having a pair of FFNs surrounding the block in the Macaron-style. Below we study these effects of these differences by mutating a Conformer block towards a Transformer block, while keeping the total number of parameters unchanged. Table~\ref{tab:ablation_model} shows the impact of each change to the Conformer block.
Among all differences, convolution sub-block is the most important feature, while having a Macaron-style FFN pair is also more effective than a single FFN of the same number of parameters.
Using swish activations led to faster convergence in the Conformer models.
\begin{table}[h]
\caption{\textbf{Disentangling Conformer.}
Starting from a Conformer block, we remove its features and move towards a vanilla Transformer block: (1) replacing SWISH with ReLU; (2) removing the convolution sub-block; (3) replacing the Macaron-style FFN pairs with a single FFN; (4) replacing self-attention with relative positional embedding~\cite{dai2019transformerxl} with a vanilla self-attention layer~\cite{vaswani2017attention}.
All ablation study results are evaluated without the external LM.}
\label{tab:ablation_model}
\begin{tabular}{@{}l|cccc@{}}
\hline 
\makecell{Model\\ Architecture} & \makecell{dev\\ clean} & \makecell{dev\\ other} & \makecell{test\\ clean} & \makecell{test\\ other} \\
\hline 
\hline 
Conformer Model & 1.9 & 4.4 & 2.1 & 4.3 \tabularnewline
\enskip -- SWISH + ReLU & 1.9 & 4.4 & 2.0 & 4.5 \tabularnewline
\enskip\enskip-- \textbf{Convolution Block} & 2.1 & 4.8 & 2.1 & 4.9 \tabularnewline
\enskip\enskip\enskip-- Macaron FFN & 2.1 & 5.1 & 2.1 & 5.0 \tabularnewline
\enskip\enskip\enskip\enskip-- {Relative Pos. Emb.} & 2.3 & 5.8 & 2.4 & 5.6 \tabularnewline
\hline 
\end{tabular}

\vskip -0.2in
\end{table}

\subsubsection{Combinations of Convolution and Transformer Modules}
\label{ablate_conformer_convolution}
We study the effects of various different ways of combining the multi-headed self-attention (MHSA) module with the convolution module. First, we try replacing the depthwise convolution in the convolution module with a lightweight convolution \cite{wu2019pay}, see a significant drop in the performance  especially on the dev-other dataset. Second, we study placing the convolution module before the MHSA module in our Conformer model and find that it degrades the results by 0.1 on dev-other. 
Another possible way of the architecture is to split the input into parallel branches of multi-headed self attention module and a convolution module with their output concatenated as suggested in \cite{wu2020lite}. We found that this worsens the performance when compared to our proposed architecture.

These results in Table~\ref{tab:sa_conv_order} suggest the advantage of placing the convolution module after the self-attention module in the Conformer block.

\begin{table}[h]

\caption{\textbf{Ablation study of Conformer Attention Convolution Blocks.} Varying the combination of the convolution block with the multi-headed self attention: (1) Conformer architecture; (2) Using Lightweight convolutions instead of depthwise convolution in the convolution block in Conformer; (3) Convolution before multi-headed self attention; (4) Convolution and MHSA in parallel with their output concatenated~\cite{wu2020lite}.
}
\label{tab:sa_conv_order}
\begin{tabular}{@{}l|cc@{}}
\hline 
\makecell{Model Architecture} & \makecell{dev\\ clean} & \makecell{dev\\ other} \\
\hline 
\hline 
Conformer & 1.9 & 4.4 \tabularnewline
{-- Depthwise conv + Lightweight convolution}  & 2.0 & 4.8 \tabularnewline
Convolution block before MHSA  & 1.9 & 4.5  \tabularnewline
Parallel MHSA and Convolution & 2.0 & 4.9 \tabularnewline
\hline 
\end{tabular}
\vskip -0.2in
\end{table}

\subsubsection{Macaron Feed Forward Modules}
\label{ablate_conformer_macaron}
Instead of a single feed-forward module (FFN) post the attention blocks as in the Transformer models, the Conformer block has a pair of macaron-like Feed forward modules sandwiching the self-attention and convolution modules. Further, the Conformer feed forward modules are used with half-step residuals.
Table~\ref{tab:ablation_macaron} shows the impact of changing the Conformer block to use a single FFN or full-step residuals.
\begin{table}[h]
\begin{centering}
\caption{\textbf{Ablation study of Macaron-net Feed Forward modules.}
Ablating the differences between the Conformer feed forward module with that of a single FFN used in Transformer models: (1) Conformer; (2) Conformer with full-step residuals in Feed forward modules; (3) replacing the Macaron-style FFN pair with a single FFN.
}
\label{tab:ablation_macaron}
\begin{tabular}{l|cccc}
\hline 
\makecell{Model\\ Architecture} & \makecell{dev\\ clean} & \makecell{dev\\ other} & \makecell{test\\ clean} & \makecell{test\\ other} \\
\hline 
\hline 
Conformer & 1.9 & 4.4 & 2.1 & 4.3 \tabularnewline
Single FFN & 1.9 & 4.5 & 2.1 & 4.5 \tabularnewline
Full step residuals & 1.9 & 4.5 & 2.1 & 4.5 \tabularnewline
\hline 
\end{tabular}
\par\end{centering}

\end{table}

\subsubsection{Number of Attention Heads}

In self-attention, each attention head learns to focus on different parts of the input, making it possible to improve predictions beyond the simple weighted average. We perform experiments to study the effect of varying the number of attention heads from $4$ to $32$ in our large model, using the same number of heads in all layers. We find that increasing attention heads up to $16$ improves the accuracy, especially over the devother datasets, as shown in Table~\ref{tab:atten_heads}.

\begin{table}[h]
\begin{centering}
\caption{Ablation study on the attention heads in multi-headed self attention.}
\label{tab:atten_heads}
\vskip -0.1in
\begin{tabular}{@{}cc|cccc@{}}
\hline 
\makecell{Attention\\ Heads} & \makecell{Dim per\\ Head} & \makecell{dev\\ clean} & \makecell{dev\\ other} & \makecell{test\\ clean} & \makecell{test\\ other} \\
\hline 
\hline 
4 & 128 & 1.9 & 4.6 & 2.0 & 4.5 \tabularnewline
8 & 64 & 1.9 & 4.4 & 2.1 & 4.3 \tabularnewline
16 & 32 & 2.0 & 4.3 & 2.2 & 4.4 \tabularnewline
32 & 16 & 1.9 & 4.4 & 2.1 & 4.5 \tabularnewline
\hline
\end{tabular}
\par\end{centering}
\vskip -0.2in
\end{table}

\subsubsection{Convolution Kernel Sizes}
To study the effect of kernel sizes in the depthwise convolution, we sweep the kernel size in $\{3, 7, 17, 32, 65\}$ of the large model, using the same kernel size for all layers. We find that the performance improves with larger kernel sizes till kernel sizes $17$ and $32$ but worsens in the case of kernel size $65$, as show in Table~\ref{tab:kernel_sizes}. On comparing the second decimal in dev WER, we find kernel size 32 to perform better than rest. 

\begin{table}[h]
\begin{centering}
\caption{Ablation study on depthwise convolution kernel sizes.}
\label{tab:kernel_sizes}
\vskip -0.1in
\begin{tabular}{@{}c|cccc@{}}
\hline 
\makecell{Kernel\\ size}  & \makecell{dev\\ clean} & \makecell{dev\\ other} & \makecell{test\\ clean} & \makecell{test\\ other} \\
\hline 
\hline 
3 & 1.88 & 4.41 & 1.99 & 4.39 \tabularnewline
7 & 1.88 & 4.30 & 2.02 & 4.44 \tabularnewline
17 & 1.87 & 4.31 & 2.04 & 4.38 \tabularnewline
32 & 1.83 & 4.30 & 2.03 & 4.29 \tabularnewline
65 & 1.89 & 4.47 & 1.98 & 4.46 \tabularnewline
\hline
\end{tabular}
\par\end{centering}
\vskip -0.2in
\end{table}

\section{Conclusion}
In this work,
we introduced Conformer, an architecture that integrates components from CNNs and Transformers for end-to-end speech recognition. We studied the importance of each component, and demonstrated that the inclusion of convolution modules is critical to the performance of the Conformer model.
The model exhibits better accuracy with fewer parameters than previous work on the LibriSpeech dataset, and achieves a new state-of-the-art performance at $1.9\%$/$3.9\%$ for test/testother.

\bibliographystyle{IEEEtran}

\bibliography{mybib}

\begin{thebibliography}{10}
\providecommand{\url}[1]{#1}
\csname url@samestyle\endcsname
\providecommand{\newblock}{\relax}
\providecommand{\bibinfo}[2]{#2}
\providecommand{\BIBentrySTDinterwordspacing}{\spaceskip=0pt\relax}
\providecommand{\BIBentryALTinterwordstretchfactor}{4}
\providecommand{\BIBentryALTinterwordspacing}{\spaceskip=\fontdimen2\font plus
\BIBentryALTinterwordstretchfactor\fontdimen3\font minus
  \fontdimen4\font\relax}
\providecommand{\BIBforeignlanguage}[2]{{%
\expandafter\ifx\csname l@#1\endcsname\relax
\typeout{** WARNING: IEEEtran.bst: No hyphenation pattern has been}%
\typeout{** loaded for the language `#1'. Using the pattern for}%
\typeout{** the default language instead.}%
\else
\language=\csname l@#1\endcsname
\fi
#2}}
\providecommand{\BIBdecl}{\relax}
\BIBdecl

\bibitem{chiu2018state}
C.-C. Chiu, T.~N. Sainath, Y.~Wu, R.~Prabhavalkar, P.~Nguyen, Z.~Chen,
  A.~Kannan, R.~J. Weiss, K.~Rao, E.~Gonina \emph{et~al.}, ``State-of-the-art
  speech recognition with sequence-to-sequence models,'' in \emph{2018 IEEE
  International Conference on Acoustics, Speech and Signal Processing
  (ICASSP)}.\hskip 1em plus 0.5em minus 0.4em\relax IEEE, 2018, pp. 4774--4778.

\bibitem{rao2017exploring}
K.~Rao, H.~Sak, and R.~Prabhavalkar, ``Exploring architectures, data and units
  for streaming end-to-end speech recognition with rnn-transducer,'' in
  \emph{2017 IEEE Automatic Speech Recognition and Understanding Workshop
  (ASRU)}.\hskip 1em plus 0.5em minus 0.4em\relax IEEE, 2017, pp. 193--199.

\bibitem{Ryan19}
Y.~He, T.~N. Sainath, R.~Prabhavalkar, I.~McGraw, R.~Alvarez, D.~Zhao,
  D.~Rybach, A.~Kannan, Y.~Wu, R.~Pang, Q.~Liang, D.~Bhatia, Y.~Shangguan,
  B.~Li, G.~Pundak, K.~C. Sim, T.~Bagby, S.-Y. Chang, K.~Rao, and
  A.~Gruenstein, ``{Streaming End-to-end Speech Recognition For Mobile
  Devices},'' in \emph{Proc. ICASSP}, 2019.

\bibitem{tara2020}
T.~N. Sainath, Y.~He, B.~Li, A.~Narayanan, R.~Pang, A.~Bruguier, S.-y. Chang,
  W.~Li, R.~Alvarez, Z.~Chen, and et~al., ``A streaming on-device end-to-end
  model surpassing server-side conventional model quality and latency,'' in
  \emph{ICASSP}, 2020.

\bibitem{graves2012sequence}
A.~Graves, ``Sequence transduction with recurrent neural networks,''
  \emph{arXiv preprint arXiv:1211.3711}, 2012.

\bibitem{vaswani2017attention}
A.~Vaswani, N.~Shazeer, N.~Parmar, J.~Uszkoreit, L.~Jones, A.~N. Gomez,
  L.~Kaiser, and I.~Polosukhin, ``Attention is all you need,'' 2017.

\bibitem{zhang2020transformer}
Q.~Zhang, H.~Lu, H.~Sak, A.~Tripathi, E.~McDermott, S.~Koo, and S.~Kumar,
  ``Transformer transducer: A streamable speech recognition model with
  transformer encoders and rnn-t loss,'' in \emph{ICASSP 2020-2020 IEEE
  International Conference on Acoustics, Speech and Signal Processing
  (ICASSP)}.\hskip 1em plus 0.5em minus 0.4em\relax IEEE, 2020, pp. 7829--7833.

\bibitem{li2019jasper}
J.~Li, V.~Lavrukhin, B.~Ginsburg, R.~Leary, O.~Kuchaiev, J.~M. Cohen,
  H.~Nguyen, and R.~T. Gadde, ``Jasper: An end-to-end convolutional neural
  acoustic model,'' \emph{arXiv preprint arXiv:1904.03288}, 2019.

\bibitem{kriman2019quartznet}
S.~Kriman, S.~Beliaev, B.~Ginsburg, J.~Huang, O.~Kuchaiev, V.~Lavrukhin,
  R.~Leary, J.~Li, and Y.~Zhang, ``Quartznet: Deep automatic speech recognition
  with 1d time-channel separable convolutions,'' \emph{arXiv preprint
  arXiv:1910.10261}, 2019.

\bibitem{han2020contextnet}
W.~Han, Z.~Zhang, Y.~Zhang, J.~Yu, C.-C. Chiu, J.~Qin, A.~Gulati, R.~Pang, and
  Y.~Wu, ``Contextnet: Improving convolutional neural networks for automatic
  speech recognition with global context,'' \emph{arXiv preprint
  arXiv:2005.03191}, 2020.

\bibitem{sainath2013deep}
T.~N. Sainath, A.-r. Mohamed, B.~Kingsbury, and B.~Ramabhadran, ``Deep
  convolutional neural networks for lvcsr,'' in \emph{2013 IEEE international
  conference on acoustics, speech and signal processing}.\hskip 1em plus 0.5em
  minus 0.4em\relax IEEE, 2013, pp. 8614--8618.

\bibitem{abdel2014convolutional}
O.~Abdel-Hamid, A.-r. Mohamed, H.~Jiang, L.~Deng, G.~Penn, and D.~Yu,
  ``Convolutional neural networks for speech recognition,'' \emph{IEEE/ACM
  Transactions on audio, speech, and language processing}, vol.~22, no.~10, pp.
  1533--1545, 2014.

\bibitem{hu2018squeeze}
J.~Hu, L.~Shen, and G.~Sun, ``Squeeze-and-excitation networks,'' in
  \emph{Proceedings of the IEEE conference on computer vision and pattern
  recognition}, 2018, pp. 7132--7141.

\bibitem{bello2019attention}
I.~Bello, B.~Zoph, A.~Vaswani, J.~Shlens, and Q.~V. Le, ``Attention augmented
  convolutional networks,'' in \emph{Proceedings of the IEEE International
  Conference on Computer Vision}, 2019, pp. 3286--3295.

\bibitem{yang2019convolutional}
B.~Yang, L.~Wang, D.~Wong, L.~S. Chao, and Z.~Tu, ``Convolutional
  self-attention networks,'' \emph{arXiv preprint arXiv:1904.03107}, 2019.

\bibitem{yu2018qanet}
A.~W. Yu, D.~Dohan, M.-T. Luong, R.~Zhao, K.~Chen, M.~Norouzi, and Q.~V. Le,
  ``Qanet: Combining local convolution with global self-attention for reading
  comprehension,'' \emph{arXiv preprint arXiv:1804.09541}, 2018.

\bibitem{wu2020lite}
Z.~Wu, Z.~Liu, J.~Lin, Y.~Lin, and S.~Han, ``Lite transformer with long-short
  range attention,'' \emph{arXiv preprint arXiv:2004.11886}, 2020.

\bibitem{lu2019understanding}
Y.~Lu, Z.~Li, D.~He, Z.~Sun, B.~Dong, T.~Qin, L.~Wang, and T.-Y. Liu,
  ``Understanding and improving transformer from a multi-particle dynamic
  system point of view,'' \emph{arXiv preprint arXiv:1906.02762}, 2019.

\bibitem{karita2019comparative}
S.~Karita, N.~Chen, T.~Hayashi, T.~Hori, H.~Inaguma, Z.~Jiang, M.~Someki,
  N.~E.~Y. Soplin, R.~Yamamoto, X.~Wang \emph{et~al.}, ``A comparative study on
  transformer vs rnn in speech applications,'' \emph{arXiv preprint
  arXiv:1909.06317}, 2019.

\bibitem{dai2019transformerxl}
Z.~Dai, Z.~Yang, Y.~Yang, J.~Carbonell, Q.~V. Le, and R.~Salakhutdinov,
  ``Transformer-xl: Attentive language models beyond a fixed-length context,''
  2019.

\bibitem{wang-etal-2019-learning-deep}
Q.~Wang, B.~Li, T.~Xiao, J.~Zhu, C.~Li, D.~F. Wong, and L.~S. Chao, ``Learning
  deep transformer models for machine translation,'' in \emph{Proceedings of
  the 57th Annual Meeting of the Association for Computational
  Linguistics}.\hskip 1em plus 0.5em minus 0.4em\relax Association for
  Computational Linguistics, Jul. 2019, pp. 1810--1822.

\bibitem{nguyen2019transformers}
T.~Q. Nguyen and J.~Salazar, ``Transformers without tears: Improving the
  normalization of self-attention,'' \emph{arXiv preprint arXiv:1910.05895},
  2019.

\bibitem{dauphin2017language}
Y.~N. Dauphin, A.~Fan, M.~Auli, and D.~Grangier, ``Language modeling with gated
  convolutional networks,'' in \emph{Proceedings of the 34th International
  Conference on Machine Learning-Volume 70}.\hskip 1em plus 0.5em minus
  0.4em\relax JMLR. org, 2017, pp. 933--941.

\bibitem{dong2018speech}
L.~Dong, S.~Xu, and B.~Xu, ``Speech-transformer: a no-recurrence
  sequence-to-sequence model for speech recognition,'' in \emph{2018 IEEE
  International Conference on Acoustics, Speech and Signal Processing
  (ICASSP)}.\hskip 1em plus 0.5em minus 0.4em\relax IEEE, 2018, pp. 5884--5888.

\bibitem{ramachandran2017searching}
P.~Ramachandran, B.~Zoph, and Q.~V. Le, ``Searching for activation functions,''
  \emph{arXiv preprint arXiv:1710.05941}, 2017.

\bibitem{panayotov2015librispeech}
V.~Panayotov, G.~Chen, D.~Povey, and S.~Khudanpur, ``Librispeech: an asr corpus
  based on public domain audio books,'' in \emph{2015 IEEE International
  Conference on Acoustics, Speech and Signal Processing (ICASSP)}.\hskip 1em
  plus 0.5em minus 0.4em\relax IEEE, 2015, pp. 5206--5210.

\bibitem{park2019specaugment}
D.~S. Park, W.~Chan, Y.~Zhang, C.-C. Chiu, B.~Zoph, E.~D. Cubuk, and Q.~V. Le,
  ``Specaugment: A simple data augmentation method for automatic speech
  recognition,'' \emph{arXiv preprint arXiv:1904.08779}, 2019.

\bibitem{largespecaugment}
D.~S. Park, Y.~Zhang, C.-C. Chiu, Y.~Chen, B.~Li, W.~Chan, Q.~V. Le, and Y.~Wu,
  ``Specaugment on large scale datasets,'' \emph{arXiv preprint
  arXiv:1912.05533}, 2019.

\bibitem{JMLR:v15:srivastava14a}
N.~Srivastava, G.~Hinton, A.~Krizhevsky, I.~Sutskever, and R.~Salakhutdinov,
  ``Dropout: A simple way to prevent neural networks from overfitting,''
  \emph{Journal of Machine Learning Research}, vol.~15, no.~56, pp. 1929--1958,
  2014.

\bibitem{jim1996analysis}
K.-C. Jim, C.~L. Giles, and B.~G. Horne, ``An analysis of noise in recurrent
  neural networks: convergence and generalization,'' \emph{IEEE Transactions on
  neural networks}, vol.~7, no.~6, pp. 1424--1438, 1996.

\bibitem{kingma2014adam}
D.~P. Kingma and J.~Ba, ``Adam: A method for stochastic optimization,''
  \emph{arXiv preprint arXiv:1412.6980}, 2014.

\bibitem{lingvo}
J.~Shen, P.~Nguyen, Y.~Wu, Z.~Chen, and et~al., ``Lingvo: a modular and
  scalable framework for sequence-to-sequence modeling,'' 2019.

\bibitem{wang2019transformer}
Y.~Wang, A.~Mohamed, D.~Le, C.~Liu, A.~Xiao, J.~Mahadeokar, H.~Huang,
  A.~Tjandra, X.~Zhang, F.~Zhang \emph{et~al.}, ``Transformer-based acoustic
  modeling for hybrid speech recognition,'' \emph{arXiv preprint
  arXiv:1910.09799}, 2019.

\bibitem{synnaeve2019endtoend}
G.~Synnaeve, Q.~Xu, J.~Kahn, T.~Likhomanenko, E.~Grave, V.~Pratap, A.~Sriram,
  V.~Liptchinsky, and R.~Collobert, ``End-to-end asr: from supervised to
  semi-supervised learning with modern architectures,'' 2019.

\bibitem{wu2019pay}
F.~Wu, A.~Fan, A.~Baevski, Y.~N. Dauphin, and M.~Auli, ``Pay less attention
  with lightweight and dynamic convolutions,'' \emph{arXiv preprint
  arXiv:1901.10430}, 2019.

\end{thebibliography}

\end{document}